\documentclass[graybox]{svmult}

\usepackage{type1cm}        
\usepackage{makeidx}         
\usepackage{graphicx}        
\usepackage{multicol}        
\usepackage[bottom]{footmisc}

\usepackage{newtxtext}       %
\usepackage{newtxmath}       

\makeindex             

\begin{document}

\title*{Introduction to stellar and substellar physics in modified gravity}
\author{Aneta Wojnar}
\institute{Departamento de F\'isica Te\'orica, Universidad Complutense de Madrid, E-28040, 
Madrid, Spain, \email{awojnar@ucm.es}
}
%
\maketitle


\abstract{We discuss the standard Lane-Emden formalism as well as the one related to the slowly rotating objects. It is preceded by a brief introduction of different forms of the polytropic equation of state. This allows to study a wide class of astrophysical objects in the framework of a given theory of gravity, as demonstrated in a few examples. We will discuss light elements burning processes and cooling models in stars and substellar objects with the use of the Lane-Emden formalism.
}

\section{Introduction}

In the spherical symmetric spacetime (hence all physical quantities are dependent on the radial coordinate $r$ only), the hydrostatic equilibrium equations are given by
\begin{align}
    \frac{d P}{d r} &= \rho\frac{d \Phi}{d r}\label{h1}\\
      \frac{dm}{dr} &= 4\pi r^2\rho(r),\label{h2}
\end{align}
where $\Phi=\Phi(r)$ is the gravitational potential, $P=P(r)$ and $\rho=\rho(r)$ are pressure and energy density, respectively, while $m=m(r)$ is mass enclosed in a spherical-symmetric ball.

In order to solve the above hydrostatic equations, we need to have boundary conditions which are given by: $\rho(0)=\rho_c$, with $\rho_c$ being the central density, $m(0)=0$, and $P(R)=0$, where $R$ is the radius of the object. Apart from them, we also need a relation between the pressure and energy density which is given by an equation of state (EoS). In the next section, we will discuss some particular forms of EoS's.

\section{Matter description in stellar interiors}

\subsection{Equation of state}\label{seceos}
As briefly discussed before, an equation of state is the crucial ingredient to solve the set of equations \eqref{h1} and \eqref{h2}. It cannot be {\it any} equation relating thermodynamics quantities such as pressure $P$ and energy density $\rho$ - it must satisfy a few particular conditions\footnote{We will focus on the barotropic EoS only, that is, when $P=P(\rho)$.} such that it can indeed describe matter inside a stellar object. One of them is the the weak energy condition $\rho>0$ and $\rho +P>0$. Apart from it, matter must not spontaneous collapse - it means, it must be microscopically stable (Le Chatelier's principle) - which provides the conditions  $P \geq 0$ and $dP/d\rho>0$. Moreover, the speed of perturbations cannot exceed the speed of light $c_s\equiv (dP/d\rho)^{1/2} \leq 1$.

There are many EoS's which fulfill those conditions and are used to study various astrophysical objects. However, we will focus now on two specific ones, given in analytical forms: ideal gas and polytropic EoS. The ideal gas is often used as an approximated description of the atmosphere, but also as one of many components of a mixture of non-interacting gases, described by the Dalton law $p_\text{total}=\sum^m_{i=1}p_i$. The ideal gas has a well known form
\begin{equation}\label{ideal}
 \rho=\frac{\mu p}{N_A k_B T}.
\end{equation}

However, we will mainly focus our attention on the polytropic form of EoS's:
\begin{equation}
P=K\rho^{1+\frac{1}{n}},
\label{Poly}
\end{equation}
where $K$ and $n$ are polytropic parameters or functions. Their exact forms depend on an astrophysical object we are interested in, and, as it will turn out, they also depend on a lot of physical processes and information which can be incorporated into their expressions. This EoS is widely used to approximate the matter description in substellar and stellar objects and also in compact stars, such white dwarfs or neutron ones. Often, it is an equation of state which one uses to analyze an astrophysical object in modified theories of gravity \cite{sotiriou2010f,2011PhR...509..167C,nojiri2011unified,nojiri2017modified}, to trace eventual problems as well as to understand the structure of equations.

Before discussing more interesting features of the polytropic EoS, let us just briefly present the simplest forms, related to the non-relativistic degenerate
electron gas for which $n=3/2$, and to relativistic one, given by $n=3$. Then, the polytropic parameters $K$'s are constants, given as \cite{stellar}
\begin{equation}\label{Ka}
    K_{n=3/2}=\frac{1}{20}\left(\frac{3}{\pi}\right)^\frac{2}{3}\frac{h^2}{m_e}\frac{1}{(\mu_e m_u)^\frac{5}{3}},\;\;\;\;K_{n=3}=\frac{hc}{8}\left(\frac{3}{\pi}\right)^\frac{1}{3}\frac{1}{m_H(\mu_e)^\frac{4}{3}},
\end{equation}
where 
\begin{equation}
 \mu_e^{-1}=X+\frac{Y}{2}+(1+X+Y)\left<\frac{Z}{A}\right>
\end{equation}
with $\left<{Z}/{A}\right>$ being the average number of electrons per nucleons in metals, $X$ and $Y$ the mass fractions of hydrogen and helium, respectively, while the other symbols have their standard meaning. Usually, one assumes $\left<{Z}/{A}\right>=1/2$ and that the star or substellar object consists of $70\%$ of hydrogen (that is, $X=0.7$). 

However, the above forms do not catch many interesting effects, such as for example more realistic description of the electron degeneracy and its time evolution, strongly coupled plasma \cite{stevenson1991search}, finite gas temperatures with phase transition points between metallic hydrogen and molecular state. It turns out that those effects can be mimicked by the polytropic EoS (or its slight modifications), which significantly simplify the calculations, especially if we do them in a framework of MG. Let us now briefly discuss some of them. In most of the cases, if not indicated, we will deal with the non-relativistic degenerate
electron gas, that is, $n=3/2$.

The degeneracy parameter $\Psi$ is defined as follows
\begin{equation}\label{degeneracy}
\Psi=\frac{\mu_{F}}{k_{B} T}=\frac{\left(3 \pi^{2} \hbar^{3}\right)^{2 / 3}}{2 m_{e} k_{B} T}\left[\frac{\rho N_{A}}{\mu_{e}}\right]^{2 / 3},
\end{equation}
where $N_A$ is the Avogadro number while the other constants have the standard meaning. Since it is dependent on the density and temperature, which change during an object's evolution, for instance, when it gravitationally contracts, its time evolution can have a non-trivial effect on the stellar and substellar properties. To take it into account, there have been a few improvements proposed such that the simple forms \eqref{Ka} acquired additional terms. For example, to cover high densities with low temperatures (degenerated gas, $\Psi>>1$) and ideal gas \eqref{ideal} ($\Psi<<1$), the polytropic constant \eqref{Ka} is \cite{burrows1993science}
\begin{eqnarray}\label{kanodeg}
 K= \frac{ (3\pi^2)^{2/3} \hbar}{5 m_e m_H^{5/3} \mu_e^{5/3}} \left( 1+ \frac{\alpha_d}{\Psi} \right),
\end{eqnarray}
with $\alpha_d\equiv5\mu_e/2\mu\approx4.82$, where $\mu$ is the mean molecular weight of ionized hydrogen/helium mixtures, and the provided value is for $X=0.75$ and $Y=0.25$. However, this EoS cannot be used for a partially degenerate gas, since $\Psi$ is considered here as a constant. Nevertheless, it has been used to get the minimum Main Sequence masses via obtaining the luminosity produced by the hydrogen burning \cite{burrows1993science,sakstein2015hydrogen,sakstein2015testing,crisostomi2019vainshtein,Olmo:2019qsj,Guerrero:2021fnz} or maximal mass of a fully convective star on the Main Sequence \cite{Wojnar:2020txr,Gomes:2022sft} in various theories of gravity. Similar modifications to the polytropic EoS are also used to study light elements' burning in fully convective stars \cite{bildsten1997lithium,ushomirsky1998light,Wojnar:2020frr}.

To consider a mixture of degenerate and ideal gas states at finite temperature as well as to take into account ionization and phase transition points, the polytropic parameter $K$ becomes a function, with the following form \cite{auddy2016analytic}
\begin{equation}\label{ka2}
K=C \mu_{e}^{-5 / 3}(1+b+ a \eta),
\end{equation}
where we have redefined the electron degeneracy as $\eta=\Psi^{-1}$. The constant $C=10^{13} \rm{cm}^4g^{-2/3}s^{-2}$ while $a=\frac{5}{2}\mu_e\mu_1^{-1}$, where $\mu_1$ is defined as
\begin{equation}\label{mu1}
\frac{1}{\mu_1}=(1+x_{H^+})X+\frac{Y}{4}.
\end{equation}
It takes into account ionization of hydrogen $X$, represented by the ionization fraction $x_{H^+}$  and depends on the phase transition points \cite{chab4}. On the other hand, the quantity $b$ reads
\begin{equation}\label{defb}
    b=-\frac{5}{16} \eta \mathrm{ln}(1+e^{-1/\eta})+\frac{15}{8}\eta^2\left( \frac{\pi^2}{3}+\mathrm{Li}_2[-e^{-1  /\eta}] \right),
\end{equation}
where $\rm{Li}_2$ denotes the second order polylogarithm function. Such an EoS is used to described matter properties, in very low-mass stars, brown dwarfs and giant exoplanets \cite{auddy2016analytic,rosyadi2019brown,Benito:2021ywe,Kozak:2022hdy}.
\begin{question}{Exercises}
    \begin{enumerate}
        \item Show that a combination of pressures $ p=p_1+p_2$, where $p_1$
        is the simple polytropic EoS with $n=3/2$ \eqref{Ka} and $p_2$ is the ideal gas one \eqref{ideal}, can be written as a polytropic EoS with the polytropic index $n=3/2$.
        \item Show that the polytropic equation of state with \eqref{ka2} in the approximation of negligible degeneracy reduces to the ideal gas.
    \end{enumerate}
\end{question}

Various forms of the polytropic EoS are also used to study terrestrial (exo-)planets. To describe a complex planets' interiors, one needs to consider a merger of the third-order finite strain Birch-Murgnagham equation of state \cite{birch1947finite} with Thomas-Fermi-Dirac one \cite{thomas1927calculation,fermi1927statistical,dirac1930note,feynman1949equations,salpeter1967theoretical}. Such a merger is well approximated by the polytropic EoS up to the pressure range $p<10^{7}\,\text{GPa}$ \cite{seager2007mass}
\begin{equation}\label{pol}
    \rho(p)=\rho_0 +cp^n,
\end{equation}
whose best-fit parameters $\rho_0$, $c$, and $n$ depend on a material the layer is composed of. The presence of $\rho_0$ in the above EoS allows to include the incompressibility of solids and liquids at low pressures. One may also make the polytropic index $n$ a variable, since it is a derivative of the inverse of
the compressibility, which is a property of the layer's material, to describe rocky and gaseous (exo-)planets \cite{weppner}.

\subsection{Other properties}
Although the equation of state is enough to solve the system of the differential equations \eqref{h1} and \eqref{h2} to get the basic properties of an astrophysical object such as mass and radius, as well as its core properties if we deal with a polytrope (see the next section), it is not sufficient to study processes happening in its interior, such us thermonuclear reactions, cooling and crystallization processes, or its evolution. Let us then discuss additional equations which are crucial for those physical problems.

If a low-mass star or a brown dwarf is massive enough\footnote{It will turn out that you will find a lot of terms as "critical mass" for something when you go deeper into some of the branches of the astrophysics.} to burn light elements in its core, the outcome of such an energy generation process is, roughly speaking, luminosity. It is obtained by the integration of the below expression:
\begin{equation}\label{Lbur}
    \frac{dL_{burning}}{dr}=4\pi r^2\dot\epsilon\rho,
\end{equation}
where the energy generation rate $\dot\epsilon$ is a function of energy density, temperature, and stellar composition. It is often approximated as a power-low function of the two first \cite{fowler}. Its (analytical) form depends on a given reaction and here we will not focus on any particular form (see section \ref{burning} for some examples). Nevertheless of the type of the reaction, such an energy is transported from the deep interior to the surface of the object. Depending on the type of the object (which can be translated as a mass criterion), one deals with the heat transport through object's interior and its atmosphere. The most common criterion
determining the class of the energy transport is provided by the Schwarzschild one
 \cite{schw,schw2}:
\begin{eqnarray}
 \nabla_{rad}&\leq&\nabla_{ad}\;\;\textrm{pure diffusive radiative or conductive transport}\\
\nabla_{rad}&>&\nabla_{ad}\;\;\textrm{ adiabatic convection is present locally,}
\end{eqnarray}
where the gradient denotes the temperature $T$ variation with depth 
\begin{equation}
 \nabla_{{rad}}:=\left(\frac{d \ln{T}}{d\ln{p}}\right)_{{rad}},
\end{equation}
and $\nabla_{ad}$ is the adiabatic temperature gradient, which in case of perfect, monatomic gas has a constant value $\nabla_{ad}=0.4$. In the case of Newtonian gravity, the criterion is given by 
\begin{equation}\label{grad}
  \nabla_{rad}=\frac{3\kappa_{rc}lp}{16\pi \bar acG mT^4},
\end{equation}
where $l$ is the local luminosity, the constant $\bar a=7.57\times 10^{-15}\frac{erg}{cm^3K^4}$ is the radiation density and $\kappa_{rc}$ is the radiative and/or conductive opacity which we will discuss later. However it was demonstrated that for MG the Schwarzschild criterion becomes modified \cite{Wojnar:2020txr} by additional terms which have a stabilizing or destabilizing effects (therefore, for instance, one can deal with more massive fully convective stars on the Main Sequence in comparison to Newtonian gravity.)

When the produced energy reaches the object's surface, it is radiated away through it. This process is well described by the Stefan-Boltzmann law ($L$ stands for luminosity)
\begin{equation}\label{stefan}
    L=4\pi f\sigma T_{eff}^4R^2,
\end{equation}
where $\sigma$ is the Stefan-Boltzmann constant and the factor $f\leq1$ which allows to include planets since they radiate less than the black-body with the same effective temperature $T_{eff}$. 

Unfortunately, the effective temperature as well as other parts of the atmospheric modelling are not only complicated to determine, but they often carry significant uncertainties. Moreover, they properties and forms also depend on the interior properties and matching conditions. However, we can keep them as parameters to be determined by some particular methods, or we can derive them from scratch. To do so, we can use for example the optical depth $\tau$, averaged over the object's atmosphere\footnote{However, it is useful for analytical studies only when we can assume that the surface gravity $g=\frac{Gm(r)}{r^2}$ is constant.} \cite{stellar,hansen}:
\begin{equation} \label{eq:od}
 \tau(r)=\bar\kappa\int_r^\infty \rho dr,
\end{equation}
where $\bar\kappa$ is a mean opacity. For objects possessing atmospheres with low temperatures, the most common mean opacity is the Rosseland one. It is given by the  Kramers' law
\begin{equation}\label{abs}
 \bar\kappa= \kappa_0 p^u T^{w},
\end{equation}
where $\kappa_0$, $u$ and $w$ are values depending on different opacity regimes \cite{planets,kley}. Another useful approximation is the assumption is the atmospheric particles satisfy the ideal gas relation \eqref{ideal}.

Another quantity which is crucial for studying processes in the stellar and substellar interiors is the thermal energy
\begin{equation}\label{Uthermal}
    U=\bar{c}_v\frac{\mathcal{M}}{A m_\mathrm{p}} T,
\end{equation}
where $T$ is the temperature of the isothermal core and $\mathcal{M}/(Am_\mathrm{p})$ is the number of ions with $A$ being the mean atomic weight. 
The mean specific heat is taken for the whole stellar configuration since it depends on the density
\begin{equation}\label{cvg}
     \bar{c}_v = \frac{1}{\mathcal M}\int_0^{\mathcal{M}} c_v(T,\rho)\frac{dm}{dr}d{r},
\end{equation}
where $c_v = c_v^\text{el}+c_v^\text{ion}$, with $c_v^\text{ion}$ being the specific heats of ions and $c_v^\text{el}$ the electrons per ions depend on density. If the main contribution comes from the ions, for which the specific heat is $c_v=(3/2)k_\mathrm{B}$, the thermal energy is reduced to the well-known form
\begin{equation} \label{thermalsimple}
    U=\frac{3}{2} k_\mathrm{B}T\frac{\mathcal{M}}{A m_\mathrm{p}}.
\end{equation}
However, if one wants to consider a more realistic case, the
dependency on the Debye temperature $\Theta_\mathrm{D}$ and the ratio of Coulomb to thermal energy  $\Gamma$ 
\begin{equation}
    \Gamma=2.28\times10^5 \frac{Z^2}{A^{1/3}} \frac{\rho_s^{1/3}}{T},
\end{equation}
with $\rho_s(T)$ being the density of the crystallized mass at a temperature $T$,
must be taken into account. They can be incorporated already in the expression for the specific heat of ions. The exact forms of the specific heats are as follows
\begin{equation}\label{cw}
      c_v^\text{el}=\frac{3}{2}\frac{k_\mathrm{B}\pi^2}{3}Z \frac{k_\mathrm{B}T}{\epsilon_\mathrm{F}},\;\;\;\;    c_v^\text{ion}=9k_\mathrm{B} \left(\frac{T}{\Theta_\mathrm{D}}\right)^3 \int_0^{\Theta_\mathrm{D}/T} \frac{x^4 e^x}{\left(e^x-1\right)^2}d{x},
\end{equation}
where $Z$ is the charge and $\epsilon_\mathrm{F}$ is the Fermi energy ($p_\mathrm{F}$ is the Fermi momentum)
\begin{align}\label{Eq: relativistic EF-pF}
\epsilon_\mathrm{F}^2 = p_\mathrm{F}^2c^2 + m_\mathrm{e}^2c^4,\;\;\;\;\;
p_\mathrm{F}^3 = \frac{3h^3}{8\pi}\frac{\rho}{\mu_\mathrm{e}m_\mathrm{p}}.
\end{align}
As evident from the form of the specific heat of ions $c_v^\text{ion}$, it depends on the crystallization properties of matter. More specifically, it depends on the critical value of the ratio of Coulomb to thermal energy $\Gamma$, denoted by $\Gamma_m$. For $\Gamma<\Gamma_m$, $c_v^\text{ion}$ reduces to $(3/2)k_\mathrm{B}$ while above this value, the specific heat of ions is given by the expression \eqref{cw}, in which  the Debye temperature is given by
\begin{equation}\label{debye}
    \Theta_\mathrm{D}=0.174\times10^4 \frac{2Z}{A}\sqrt{\rho}.
\end{equation}
We see that all those quantities depend on density, which is obtained from the hydrostatic equilibrium equation. Because of that fact, depending on the gravity framework one works, those matter properties do depend on it. As we will see in the section \ref{sec}, it has an important consequence on the cooling and crystallization processes in white dwarfs.

\section{Lane-Emden formalism}

Although the Lane-Emden formalism is well-known and it can be find in almost every standard textbook, let us briefly recall it before moving to its MG counterpart. Roughly speaking, it is a dimensionless Poisson equation with the hydrostatic equilibrium equation and polytropic form of the equation of state in the spherical-symmetric spacetime. As we will see in the further part, it is widely used in many astrophysical problems.

\subsection{Newtonian case}\label{newton}

Let us consider the Poisson equation
\begin{equation}
\nabla^2\Phi = -4\pi G \rho,
\label{Poisson0}
\end{equation}
where $\Phi$ is the gravitational potential, $\rho$ is energy density while $G$ the Newton's gravitational constant. Considering spherical-symmetric spacetime such that all variables are the functions of the radial coordinates, that is, $\Phi\equiv\Phi(r)$ and $\rho\equiv\rho(r)$, the Poisson equation \eqref{Poisson0} and corresponding hydrostatic equilibrium equation take the following form
\begin{equation}
\frac{1}{r^2}\frac{d}{d r}\Big(r^2 \frac{d \Phi}{d r}\Big) =-4\pi G \rho,\;\;\;\;\;\frac{d P}{d r} = \rho\frac{d \Phi}{d r}, \label{pair}
\end{equation}
where $P\equiv P(r)$. Applying to both the polytropic equation of state \eqref{Poly}, these equations can be written as one
\begin{equation}
\frac{1}{\xi^2}\frac{d}{d \xi}\Big(\xi^2 \frac{d \theta}{d \xi}\Big) =-\theta^n,
\label{LE}
\end{equation}
where $\theta$ is a function of $\xi$. It satisfies the boundary conditions $\theta(0)=1$, $\theta'(0)=0$, where the prime $'\equiv \frac{d}{d\xi}$. The above equation is the Lane-Emden equation (LEE). The relations between the dimensionless variables $\theta$ and $\xi$ and energy density $\rho$ and radial coordinate $r$ are like follows:
\begin{equation}
\rho=\rho_c \theta^n, ~~~~~ r=r_c\xi ~~~~~{\rm with}~~~~~ r_c^2=\frac{K(n+1)\rho_c^{(\frac{1}{n}-1)}}{4\pi G}.
\label{nondimensional0}
\end{equation}
where $\rho_c$ denotes the central density. Notice that from \eqref{Poly} we have
\begin{equation}
    P=P_c \theta^{n+1},
\end{equation}
where $P_c:=K\rho_c^\frac{n+1}{n}$.
\begin{question}{Exercises}
\begin{enumerate}
    \item Derive the Lane-Emden equation \eqref{LE}.
    \item It turns out that the LEE possesses three exact solutions for some specific values of the polytropic index $n$. Find the exact solutions of the LEE \eqref{LE} for $n=0$, $n=1$, and $n=5$.
    \item In some particular cases, such as for instance studying processes happening in the stellar core (see the section \ref{burning}), the near-center solution is needed and quite easy to find with respect to the the general one. Therefore, find the approximated solution of \eqref{LE} around $\xi\approx0$, that is, an analytic form of $\theta(\xi\approx0)$.
\end{enumerate}
\end{question}
Knowing the solution $\theta$ of the LEE for the chosen polytropic index $n$ and polytropic (constant or variable) $K$, one can immediately get the characteristics of a given astrophysical object. Therefore, the total stellar mass $M$ and its radius $R$ are given as
\begin{equation}\label{mr}
     M=4\pi r_c^3\rho_c\omega_n,\;\;\;
 R=\gamma_n\left(\frac{K}{G}\right)^\frac{n}{3-n}M^\frac{1-n}{n-3},
\end{equation}
where 
\begin{equation}
     \omega_n=-\xi_1^2\frac{d\theta}{d\xi}\Big|_{\xi=\xi_1},\;\;\;\;\;
  \gamma_n=(4\pi)^\frac{1}{n-3}(n+1)^\frac{n}{3-n}\omega_n^\frac{n-1}{3-n}\xi_1
\end{equation}
and $\xi_R$ is a value for which $\theta(\xi_R)=0$. Thus, the first zero of $\theta$ indicates the radius of the object.
\begin{question}{Exercise}
Derive the mass and radius given by \eqref{mr}.
\end{question}
 The density and pressure profiles are provided by \eqref{nondimensional0} and \eqref{Poly}, respectively, while the temperature profile yields
\begin{equation}\label{temprof}
     T=K \frac{m_H\mu }{k_B}\rho_c^\frac{1}{n}\theta,
\end{equation}
with $k_B$ being Boltzmann's constant, $m_H$ the mass of Hydrogen atom, and $\mu$ the mean molecular weigh.
\begin{question}{Exercise}
Derive the temperature profile \eqref{temprof}.
\end{question}

Moreover, one can also derive the core quantities, such as the core temperature $T_c$, core density $\rho_c$, and core pressure $P_c$:
\begin{equation}
 \rho_c=\delta_n\left(\frac{3M}{4\pi R^3}\right),\;\;\;
 T_c=K \frac{m_H\mu }{k_B}\rho_c^\frac{1}{n},\;\;\;
 P_c:=K\rho_c^\frac{n+1}{n}
\end{equation}
 where $
 \delta_n=-\frac{\xi_1}{3\frac{d\theta}{d\xi}|_{\xi=\xi_1}}. $

\begin{question}{Exercise}
Find mass, radius, central temperature and density for a polytropic star with respect to the solutions of the Lane-Emden equation for $n=1.5$ and $n=3$. 
\end{question}

To know more about the LEE, its generalization and further applications, see \cite{horedt2004polytropes}. Now on, we will focus on its MG form.

\subsection{Modified gravity}

If you are interested in modified theories of gravity, you already now that some of those proposal can also modify the Newtonian limit equations. Therefore, the Poisson and hydrostatic equilibrium equations acquire additional terms whose forms depend on a particular theory \cite{banados2010eddington,koyama2015astrophysical,Olmo:2019flu,toniato2020palatini,olmo2021parameterized}. We can then write the first equation in a generic way for the spherical-symmetric spacetime as 
\begin{equation}
\frac{1}{r^2}\frac{d}{d r}\Big(r^2 \frac{d \Phi}{d r}\Big) = -4\pi G \rho +\text{mgt}(r),
\label{Poisson1}
\end{equation}
where the modified gravity term $\text{mgt}(r)$ is a general function, characteristic for a given theory of gravity. Here we also assume that all additional elements of that function are $r-$coordinate dependent.

Analogously, employing the polytropic EoS \eqref{Poly} and the hydrostatic equilibrium equation \eqref{h1}, we can rewrite it as the modified Lane-Emden equation (MLEE)
\begin{equation}
\frac{1}{\xi^2}\frac{d}{d \xi}\Big(\xi^2 \frac{d \theta}{d \xi}\Big) =-\theta^n +  g_{mod0}(\xi)
\label{modLEEn}
\end{equation}
where
\begin{equation}\label{gmod}
g_{mod0}=\frac{\text{mgt}(r)}{4\pi G \rho_c}
\end{equation}
is a dimensionless term induced by a given theory. Notice its dependence on $\rho_c$. As you will see in the following examples, usually the core density can be hidden into a re-scaled theory parameter.
\begin{question}{Exercise}
Derive the MLEE \eqref{modLEEn}.
\end{question}

\begin{example}{Examples: Modified Lane-Emden equations}
\begin{itemize}
    \item Scalar-tensor theories (Horndeski and beyond,...) \cite{broken,sak1,sak2}
    \begin{equation} 
    \frac{1}{\xi^{2}} \frac{\mathrm{d}}{\mathrm{d} \xi}\left[\left(1+\frac{n}{4} \Upsilon \xi^{2} \theta^{n-1}\right) \xi^{2} \frac{\mathrm{d} \theta}{\mathrm{d} \xi}+\frac{\Upsilon}{2} \xi^{3} \theta^{n}\right]=-\theta^{n},
\end{equation}
where $\Upsilon$ in the theory parameter.
    \item Eddington-inspired Born Infeld gravity (EiBI) \cite{Pani:2011mg}
    \begin{equation}
\frac{d}{d\xi}\left(  \xi^2  \, \frac{d\theta}{d\xi} \left[ 1 +\alpha\theta^{n-1}   \right] \right)= - \xi^2 \theta^n \ ,
\end{equation}
where the EiBI corrections are hidden in the dimensionless parameter $\alpha=\frac{ \epsilon \, n}{2 \, r_c^2}$.
Notice that it depends on the polytropic parameter and also on the star's central energy density $\rho_c$.
    \item Palatini $f(R)$ gravity
\cite{aneta1}
\begin{equation}\label{LE}
 \frac{1}{\xi}\frac{d^2}{d\xi^2}\left[\sqrt{\phi}\xi\left(\theta-\frac{2\alpha}{n+1}\theta^{n+1}\right)\right]=
 -\frac{(\phi+\frac{1}{2}\xi\frac{d\phi}{d\xi})^2}{\sqrt{\phi}}\theta^n,
\end{equation}
where $\phi=1+2\alpha \theta^n$ while $\alpha=\kappa c^2\beta\rho_c$ is the rescaled modifed gravity parameter.
    \item Metric $f(R)$ gravity has a much more complex form, see \cite{capLE}.
\end{itemize}
\end{example}

\subsection{Slowly rotating astrophysical objects}
The formalism presented above turn out to be very useful (see the section \ref{appl}) in studying non-relativistic objects in modified gravity. However, everything in the Universe rotates, therefore if one day you would like to consider a more realistic models to, for example, test your model of gravity (or other fundamental interactions) against observational data, the rotation\footnote{As well as magnetic field, energy transport properties, time dependency and many others...} should be included.

Can we however have a similar formalism in modified gravity but for the rotating polytropes? The answer is yes - we can adopt the derivation presented in \cite{chandrasekhar1933equilibrium} and write down the MLEE for a rotating object, as it was done in \cite{CBW}. Moreover, it turns out that one can even provide a generic solution of such an equation. Let us see how it can be done.

Let us consider then a slowly rotating object along the $z$-axis with the uniform angular speed $\omega$. Adopting the polar coordinates \{$r, \mu(=\cos\vartheta), \phi$\}, the equations of hydrostatic equilibrium in that case are written as
\begin{align}
\frac{\partial P}{\partial r} = \rho\frac{\partial \Phi}{\partial r} + \rho\omega^2r(1-\mu^2)~,\;\;\;\;\;
\frac{\partial P}{\partial \mu} = \rho\frac{\partial \Phi}{\partial \mu} - \rho\omega^2r^2\mu
\end{align}
in which we have neglected $\phi$ by assuming the axial symmetry. The gravitational potential $\Phi$, as previously, carries additional terms introduced by the given theory of gravity. Moreover, the extra term in the Poisson equation \eqref{Poisson1} also gains the $\mu-$dependence due to the rotation, that is:
\begin{equation}
\nabla^2\Phi = -4\pi G \rho +\text{mgt}(r,\mu),
\label{Poisson}
\end{equation}
where, analogously to the non-rotating case, the term $\text{mgt}(r,\mu)$ is a general function. Apart from it, the energy density also depends now on the angular coordinate $\mu$, that is, $\rho=\rho(r,\mu)$ as well as pressure, such that the polytropic EoS is now
\begin{equation}
P(r,\mu)=K\rho^{1+\frac{1}{n}}.
\label{Polytrope}
\end{equation} 
Introducing the dimensionless variables, we also need to remember about the $\mu-$dependence, thus
\begin{equation}
\rho=\rho_c \Theta^n, ~~~ r=r_c\xi ~~~{\rm with}~~~ r_c^2=\frac{K(n+1)\rho_c^{(\frac{1}{n}-1)}}{4\pi G} 
\label{nondimensional}
\end{equation}
where $\Theta\equiv(\xi,\mu)$ is a function of both $\xi$ and $\mu$. Then, we can derive the MLEE for the rotating polytrope of the form:
\begin{align}
\frac{1}{\xi^2}\frac{\partial}{\partial \xi}\Big(\xi^2 \frac{\partial \Theta}{\partial \xi}\Big) + \frac{1}{\xi^2}\frac{\partial}{\partial \mu}\Big((1-\mu^2) \frac{\partial \Theta}{\partial \mu}\Big) 
= v + g_{mod}(\xi,\mu)-\Theta^n
\label{modLEE}
\end{align}
where we have introduced a dimensional parameter $v=\omega^2/2\pi G \rho_c$. It can be interpreted as a measure of the outward centrifugal force compared to the self-gravity of the rotating polytrope. On the other hand, $g_{mod}(\xi,\mu)=\text{mgt}(r,\mu)/4\pi G \rho_c$ is the dimensionless modification term whose form is explicit when a gravitational proposal is chosen. 
\begin{question}{Exercise}
    Derive the MLEE for the rotating polytrope \eqref{modLEE}.
\end{question}

As already mentioned, one can find an exact solution $\Theta(\xi,\mu)$ of \eqref{modLEE} in the form of the solutions of the non-rotating MLEE which is usually much easier to solve than its rotating counterpart. It turns out that such a solution for a slow rotation has the following form \cite{CBW}:
\begin{align}\label{solrot}
\Theta(\xi,\mu)=\theta(\xi) 
+ v\Big[\psi_0(\xi) + A_2\psi_2(\xi)P_2(\mu)\Big],
\end{align}
where $P_2(\mu)$ is the Legendre function while the quantity $A_2$ is given by
\begin{equation}
A_2= - \frac{5}{6}\frac{\xi_R^2}{[3\psi_2(\xi_R)+\xi_R \psi_2^{'}(\xi_R)]}.
\label{A2}
\end{equation}
The dimensionless radius $\xi_1$ is the first zero of $\theta(\xi)$ and $'$ denotes derivative with respect to $\xi$. The functions $\psi_0$ and $\psi_2$ satisfy particular differential equations which depend on modified gravity terms. You may check how they look like in \cite{CBW}. The derivation of \eqref{solrot} is quite tedious and technical, and we will skip it here. 

Since the differential equations the function $\psi_0$ and $\psi_2$ need to satisfy depend on a model of gravity, it can happen that one cannot solve those equations to get their exact form. However, it was shown \cite{CBW} that for some models of gravity such as
\begin{example}{Examples: Modified Poisson equations}
  \begin{itemize}
    \item Scalar-tensor theories (Horndeski and beyond,...)
    \begin{equation}
	\nabla^2 \Phi \sim -\frac{\kappa}{2}\Big(\rho+\frac{\Upsilon}{4}\nabla^2(r^2\rho)\Big)
	\end{equation}
    \item Eddington-inspired Born Infeld gravity 
    \begin{equation}
	\nabla^2 \Phi \sim -\frac{\kappa}{2}\Big(\rho+\frac{\epsilon}{2}\nabla^2\rho\Big)
	\end{equation}
    \item Palatini $f(R)$ gravity \cite{toniato2020palatini}
	\begin{equation}
	\nabla^2 V \sim -\frac{\kappa}{2}\Big(\rho+2\beta\nabla^2\rho\Big)
	\end{equation}
\end{itemize}
\end{example} 
one can indeed solve them. It seems that the formalism presented in that section can be applied to any theory of gravity of which the additional term appearing in the Poisson equation is a function of density, its higher order radial-derivatives, or its Laplacian \cite{CBW}.

\section{Applications}\label{appl}
Since an application of the introduced formalism (and its extension to, for instant, even more general EoS) is plentiful, we will briefly review the existing literature and have a closer look on three astrophysical processes in which the LE formalism has a direct use. Therefore, we will focus only on some parts of the stellar and substellar evolution, with the main focus on the objects
with masses not exceeding $0.5M_\odot$, where $M_\odot$ is the solar mass. The reason for such a limitation is that the interiors of such objects, even if they are true stars on the Main Sequence, are fully convective. This means that their interior properties are well described by the polytropic EoS which we have discussed at the beginning of that chapter, while the LE formalism presented afterwards allows us to obtain necessary ingredients to study non-trivial problems in astrophysics. 

Roughly speaking, before reaching the Main Sequence, a stellar object, called pre-Main Sequence (PMS) star, is still contracting. On the HR diagram it follows the so-called Hayashi track and it decreases its luminosity but does not change too much its surface temperature. Depending on the core conditions (mainly on its temperature, but also, as it turns out, a model of gravity), the PMS star can already  burn light elements such as deuterium and lithium. The temperature of the hydrogen ignition, on the other hand, is much higher than that of the other light elements. Moreover, when a PMS star starts burning this element stably, it becomes a true star and enters the Main Sequence phase of the stellar evolution. The energy generated in the core induces pressure which balances the gravitational pull such that the star stops contracting.  However, if the central temperature and other core conditions are not enough to enable the stable hydrogen burning, the object further contracts till the electron degeneracy pressure is high enough to counterbalance the gravitational attraction. Such an object, called a brown dwarf, radiates its energy away, and since it does not possess any other energy source apart from the one gained during its contraction phase, it cools down with time. As we will see, all those process are gravity model dependent. Before however doing it, let us briefly mention other works in which polytropes and the LE formalism were or can be used.

Since in modified gravity the hydrostatic equilibrium equations \cite{saito2015modified,kozak2021invariant,olmo2021parameterized} (for review, see~\cite{Olmo:2019flu}) and matter description \cite{wojnarfermi} are modified, the changes in the astrophysical object's internal properties and in its evolution are also expected to happen. Because of that fact, a few tests with the use of stars, brown dwarfs, (exo-)planets and white dwarfs have been proposed. The most common feature of those objects which is affected by these modifications are limited masses, such as the Chandrasekhar mass-limit of WDs~\cite{1935MNRAS..95..207C,Saltas2018white,jain2016white,banerjee2017constraints,wojnar2021white,belfaqih2021white,2022PhRvD.105b4028S,2022PhLB..82736942K,2021ApJ...909...65K}, the minimum Main Sequence mass~\cite{sakstein2015hydrogen,sakstein2015testing,crisostomi2019vainshtein,Olmo:2019qsj}, minimum mass for deuterium burning~\cite{rosyadi2019brown}, Jeans~\cite{capozziello2012jeans} and opacity mass~\cite{wojnar2021jupiter}.
Another tool which is used to constrain different theories of gravity are seismic data obtained from the helioseismology \cite{saltas2019obtaining,saltas2022searching} or seismic analysis from the terrestrial planets \cite{Kozak:2021ghd,Kozak:2021zva,Kozak:2021fjy}. It turns out that also the light elements' abundances in stellar atmospheres~\cite{Wojnar:2020frr} are affected when the gravitational framework is different from the Newtonian one. The evolutionary phases of such objects as non-relativistic stars~\cite{Wojnar:2020txr, chowdhury2021modified,Guerrero:2021fnz, straight2020modified,Gomes:2022sft,chowdhury2022revisiting}, brown dwarfs~\cite{Benito:2021ywe,Kozak:2022hdy}, WDs~\cite{2022PhRvD.105b4028S,Kalita:2022zki}, and giant planets~\cite{wojnar2021jupiter,Wojnar:2022ttc} considered in modified gravity also differ with respect to their Newtonian description. Moreover, those objects can be used to constrain different theories when more accurate data provided by GAIA, James Webb Space Telescope, or Nancy Grace Roman Space Telescope are available.

\subsection{Light elements' burning}\label{burning}
As an example of the use of the introduced formalism, let us discuss two important thermonuclear processes which happen in the PMS and stellar interiors. To examine them, we will need solutions of the LE equation. Since they depend on a model of gravity, the results one will obtain will differ with respect to Newtonian ones. As we will also discuss later on, this conclusion carries important consequences and possibilities.

\subsubsection{Lithium}\label{lithium}
Since we will deal with a low-mass PMS star with mass $M$ as already mentioned, it is fully convective (apart from the radiative atmosphere). Because of that fact we can safely assume that it is well-mixed, so we do not need to worry about the non-homogeneous distribution of various elements. It also means that the matter properties are well-described by the polytropic EoS and the LE formalism. Denoting the hydrogen fraction as $X$,
the depletion rate of $^7\textrm{Li}$ is expressed by
\begin{equation}\label{reac}
 M\frac{{d}f}{{d}t}=-\frac{Xf}{m_H}\int^M_0\rho\langle\sigma v\rangle dM,
\end{equation}
where $f$ is the lithium-to-hydrogen ratio. Since we deal with a non-resonant reaction rate in the temperature range $T<6\times 10^6$ K, it is given by
\begin{equation}
 N_A\langle\sigma v\rangle=Sf_{scr} T^{-2/3}_{c6}\textrm{exp}\left[-aT_{c6}^{-\frac{1}{3}}\right]\;\frac{\textrm{cm}^3}{\textrm{s g}},
\end{equation}
where $T_{c6}\equiv T_c/10^6$ K is the core temperature and $f_{scr}$ is the screening correction factor. The dimensionless parameters $S=7.2\times10^{10}$ and $a=84.72$ were fitted to the reaction rate $^7\textrm{Li}(p,\alpha)\,^4\textrm{He}$ \cite{ushomirsky1998light,cf,raimann}.

Immediately we see from the above formulas that in order to calculate the lithium depletion rate, we need to find out the central temperature $T_c$ and density $\rho$. Fortunately, from the section \ref{newton} and for a given theory of gravity we can write that the core characteristic are given by
\begin{align}\label{ctemp}
 T_c=&1.15\times 10^6 \left(\frac{\mu_\text{eff}}{0.6}\right)\left(\frac{M}{0.1M_\odot}\right) \left(\frac{R_\odot}{R}\right)
 \frac{\delta^\frac{2}{3}}{\xi_R^\frac{5}{3}(-\theta'(\xi_R))^\frac{1}{3}}\text{K}\\
 \rho_c=&0.141\left(\frac{M}{0.1M_\odot}\right) \left(\frac{R_\odot}{R}\right)^3\delta\,\frac{\text{g}}{\text{cm}^3},
\end{align}
where the mean molecular weight $\mu_\text{eff}$ with the electron degeneracy taken into account is given by the expression
\begin{equation}
    \frac{1}{\mu_\text{eff}}=\frac{1}{\mu_i}+\frac{1}{\mu_e}\frac{2F_{3/2}(\eta)}{3 F_{1/2}(\eta)}
\end{equation}
in which $\mu_i=\rho N_A/n_i$ is the mean molecular weight of the gas, $\eta$ the electron degeneracy and $F_n(\eta)$ the $n$th order Fermi-Dirac function. We see that the values of the core quantities are MG dependent via the solutions of the modified LE, represented here by $\delta$, $\xi_R$, and $\theta'$.

 For such a case, the radius of the PMS star is given by \cite{ushomirsky1998light} 
\begin{equation}
 \frac{R}{R_\odot}\approx\frac{7.1\times10^{-2}\gamma}{\mu_\text{eff}\mu_e^\frac{2}{3}F^\frac{2}{3}_{1/2}(\eta)}
 \left(\frac{0.1M_\odot}{M}\right)^\frac{1}{3}\label{Rpol}.
\end{equation}
\begin{question}{Exercises}
\begin{enumerate}
    \item  Find the polytropic function $K$.
    \item Change the integration variable to the spatial ones in \eqref{reac}.
    \item Use the LE formalism to rewrite \eqref{reac} in the form of the LE variables.
\end{enumerate}
\end{question}
Following the steps given in the above exercises, the depletion rate of lithium will have the following form ($u\equiv aT_{c6}^{-1/3}$):
\begin{align}\label{rate}
  \frac{\text{d}}{\text{d}t}\text{ln}f&\sim\left(\frac{u}{a}\right)^2\int_0^{\xi_R}f_\text{scr} h(\theta,u)d\xi\,\,\,\,  \frac{1}{\text{s}},
\end{align}
where $ h(\theta,u)$ is a function of $u$ and the solution of the LE equation, to be determined in the above exercise. It is very improbable that we will be able to solve the modified LE equation with $n=3/2$ analytically, and to be able to go forward with our calculations without the need of numerical methods. However, we should notice that the burning processes in low-mass stars happen in the star's core, therefore the near-center approximation $\theta(\xi\approx0)$, as obtained in one of the exercise in the section \ref{newton} is sufficient for our purposes. Depending on the theory of gravity, its form can also depend on the theory parameter.

Integration in the equation \eqref{rate} can be easily performed with the assumption that the time evolution of the degeneracy parameter $\eta$ is not significant with respect to the changes in the star's size (that is, the radius). Therefore, for the PMS stars with masses $M>0.2M_\odot$ the lithium depletion $\mathcal{F}$ is
\begin{eqnarray}\label{ratio}
       \mathcal{F} &=& \ln \frac{f_0}{f} = 5.6 \times 10^{14} T_{eff}^{-4}  \left(\frac{X}{0.7}\right)\left(\frac{0.1 M_\odot}{M} \right)^3 \left(\frac{0.6}{\mu_{eff}} \right)^6 \\
       &\times&
    S f_{scr}  a^{16}g(u) \Big(1 + \text{eventual MG term} \Big) \frac{\xi_R^7(-\theta'(\xi_R))^2}{\delta^2},\nonumber
\end{eqnarray}
where $f_0$ is the initial abundance, $ T_{eff}$ the effective temperature,
 $g(u)= u^{-37/2}e^{-u}-29\Gamma(-37/2,u)$, and $\Gamma (-37/2,u)$ is the upper incomplete gamma function. The effect of MG is clearly visible, even without the eventual MG modification term, since the last fraction includes the modified LE solutions.
\begin{question}{Exercises}
    \begin{enumerate}
        \item Find the radius and luminosity as functions of time using the Stefan-Boltzmann equation and the virial theorem.
        \item Write down the contraction time: 
\begin{eqnarray}\label{cont}
    t_{cont} &=& -\frac{R}{dR/dt} 
\end{eqnarray}
in terms of the central temperature $T_c$.
\item Assuming that the PMS star depletes its lithium when it reaches the Main Sequence, such that the contraction time $t_{cont}$ is comparable to the destruction time $t_{dest}$, find the central temperature, age, radius, and luminosity of a $0.5M_\odot$ star. The destruction time is given by
\begin{eqnarray}\label{destr}
       t_{dest} &=& \frac{m_P}{X \rho <\sigma v>}. 
\end{eqnarray}
    \end{enumerate}
\end{question}

\subsubsection{Hydrogen}\label{Hydro}
When the young star reaches the Main Sequence, it means that in its core the temperature is high enough to start the hydrogen ignition. The energy generated by this process is radiated away through the star's surface. It turns out that this stable process happens for sufficiently massive objects. This critical mass, called minimum Main Sequence mass, we are going to calculate now.

The energy generation rate per unit mass for the hydrogen ignition process has the power law form \cite{fowler,burrows1993science}
\begin{equation} \label{eq:pp}
\dot{\epsilon}_{pp}= \dot{\epsilon}_c \left(\frac{T}{T_c}\right)^s \left(\frac{\rho}{\rho_c} \right)^{u-1}, \,\,\,\,\,\dot{\epsilon}_c=\epsilon_0T_c^s\rho_c^{u-1} ,
\end{equation}
where the two exponents can be approximated as  $s \approx 6.31$ and $u \approx 2.28$, while $\dot{\epsilon}_0\approx 3.4\times10^{-9}$ ergs g$^{-1}$s$^{-1}$. Considering again a PMS star with the hydrogen fraction $X=0.75$, the number of baryons per electron in low-mass stars is $\mu_e \approx 1.143$. The luminosity being an effect of the hydrogen burning reads
\begin{equation}\label{Eq: Luminosity}
L_{HB}=\int \dot{\epsilon}_{pp}\, dM,
\end{equation}
in which we can again use the LE formalism. Considering the polytrope with the polytopic function $K$ given by \eqref{kanodeg} and the near-zero solution $\theta(\xi\approx0)$, we find that
 \begin{equation}\label{lhb}
  \frac{L_{HB}}{L_\odot}=
\frac{1.53\times10^7\Psi^{10.15}}{(\Psi+\alpha_d)^{16.46}}
\frac{\delta^{5.487}_{3/2}M^{11.977}_{-1}}{\omega_{3/2}\gamma^{16.46}_{3/2}},
 \end{equation}
where $L_\odot$ is the solar luminosity while $M_{-1}=M/(0.1M_{\odot})$. As previously, the MG gravity effect is present via the solutions of the modified LE equation.
\begin{question}{Exercise}
    Following the similar steps as in the subsection \ref{lithium}, derive \eqref{lhb}.
\end{question}
To find the minimum Main Sequence mass, we need to write down the Stefan-Boltzmann equation \eqref{stefan} (with $f=1$ for the black body approximation) as a function of mass $M_{-1}$. Equaling this two luminosities, that is, $L_{HB}=L$ will provide the critical mass $M_\text{HB}$ which a PMS star needs to have in order to ignite hydrogen and burn it in a stable way.

From the LE formalism we have already the radius written as a function of mass \eqref{mr} but the effective temperature, which we assume to be the temperature of the photosphere from which the energy is radiated away, is much more complicated to be obtained. Usually, one uses the matching procedure of the specific entropies calculated in the stellar interior and in the photosphere \cite{burrows1993science}
\begin{equation}\label{Eq: photospheric temperature}
T_{ph}=\frac{1.8 \times 10^6 \rho _{ph}^{0.42}}{\Psi ^{1.545}} \ .
\end{equation}
\begin{question}{Exercises}
Using the definition of the photopshere (for which the optical depth \eqref{eq:od} is equaled to $2/3$) and the assumption on the constant surface gravity
    \begin{enumerate}
        \item Find the photosheric pressure $p_{eff}$.
        \item Assuming that the photosphere matter can be approximated by the ideal gas, use \eqref{Eq: photospheric temperature} and obtained before $p_{eff}$ to find the photospheric density $\rho_{eff}$.
        \item Write down the Stefan-Boltzmann law as a function of mass $M_{-1}$.
        \item Show that the minimum Main Sequence mass is given by
        \begin{equation} \label{result}
M_{-1}^{MMSM}=0.290 \frac{\gamma_{3/2}^{1.32} \omega_{3/2}^{0.09}}{\delta_{3/2}^{0.51}} \frac{(\alpha_d + \Psi)^{1.509}}{\Psi^{1.325}} \left(1+\text{eventual MG term}\right)
\end{equation}
    \end{enumerate}
\end{question}

\subsection{Cooling processes}
\subsubsection{Brown dwarfs}
Since the conditions occurring in the core are not enough to ignite hydrogen and to produce radiation pressure to balance the gravitational pull, PMS objects (from now on called brown dwarfs) with masses lower than $M_{-1}^{MMSM}$ continue contracting. There is no stable energy production, thus a brown dwarf radiates the stored energy away and cools down with time. Because of the ongoing contraction, the degenerate gas' contribution starts being relevant and moreover, it changes with time as the brown dwarf is still contracting before reaching the equilibrium, that is, when the electron degeneracy pressure balances the gravitational one. Therefore, we can still deal with the polytrope, however with the function $K$ adjusted to the mixture of degenerate and ideal gas states at finite temperature (and other improvements, see the section \ref{seceos}), given by \eqref{ka2}. Similarly as in the case of low-mass stars, the biggest challenge is to derive the effective temperature which appears in the Stefan-Boltzmann equation \eqref{stefan}. Again using the entropy matching procedure, 
the effective temperature $T_\text{eff}$ is obtained in terms of the degeneracy parameter $\eta(=\Psi^{-1})$ and the photospheric density $\rho_\text{ph}$ as \cite{auddy2016analytic}
\begin{equation}\label{tsur}
    T_\text{eff}=b_1 \times 10^6 \rho_\text{ph}^{0.4}\eta^\nu\,\,\text{K},
\end{equation}
where the parameters $b_1$ and $\nu$ takes different values since they depend on the specific model adopted for describing the phase transition between a metallic hydrogen and helium state (the brown dwarf's interior) and the photosphere (composed of molecular hydrogen and helium).
\begin{question}{Exercise}
    Following the analogous steps from the section \ref{Hydro}, derived the effective temperature and show that the luminosity is given by 
    \begin{equation}\label{lumph}
    L=\frac{0.0721 L_\odot}{\kappa_R^{1.1424}\gamma^{0.286}}
    \left(\frac{M}{M_\odot} \right)^{1.239}
    \frac{\eta^{2.856\nu}b_1^{2.856}}{(1+b+a\eta)^{0.2848}}
    \left(1+\text{eventual MG term}\right).
\end{equation}
\end{question}
However, the electron degeneracy $\eta$ is a function of time, hence we need to find that dependence. This can be done again with the interior entropy (which is a function of $\eta$) and the thermodynamics laws. The result of this procedure yields
    \begin{eqnarray}\label{eta}
    \frac{\rm d\eta}{\rm dt}=&-&
    \frac{1.1634\times10^{-18}b_1^{2.856}\mu_{1{mod}}}{\kappa_R^{1.1424}\mu_e^{8/3}}
    \left(\frac{M_\odot}{M} \right)^{1.094}
    \\
    &\times&\eta^{2.856\nu}(1+b+a\eta)^{1.715} \gamma^{0.7143} \left(1+\text{eventual MG term} \right),\nonumber
\end{eqnarray}
where $
 \frac{1}{\mu_{1mod}}=\frac{1}{\mu_1}+\frac{3}{2}\frac{x_{H^+}(1-x_{H^+})}{2-x_{H^+}}$ and $\mu_1$ is given by \eqref{mu1}.
The equations \eqref{eta} and (\ref{lumph}) with the initial conditions $\eta=1$ at $t=0$, provides the model of cooling process for a brown dwarf star in a given theory of gravity.

\subsubsection{White dwarfs}\label{sec}
The final state of a Main Sequence star with mass $\lesssim (10\pm2)\, M_\odot$ is a white dwarf - a very compact star which mainly consists of a core of the progenitor star. The white dwarf's mass ranges from a bit above of the solar mass, but with a much smaller radius. Since in that stage of the stellar evolution the main energy source is the energy stored during the previous active phases (that is, when the thermonuclear reactions were taking place in the stellar interior), such a stellar remnant also cools down with time. Here, we will focus on the process of cooling in which we take into account crystallization. When the white dwarf's core starts crystallizing, latent heat is released, contributing the thermal energy which is radiated away. This delays the cooling process. As soon as the star solidifies, the specific heat follows the Debye law (known as Debye cooling). The simplest model of cooling in which one takes into account only the radiation of the total thermal energy of the star \eqref{thermalsimple} in modified gravity was given in \cite{Kalita:2022zki}. In what follows, we will demonstrate how to include dependence on the Debye temperature and the ratio of Coulomb to thermal energy to it, and we will also insert the crystallization process.

Therefore, the luminosity resulting from a decrease in thermal energy \eqref{Uthermal} of ions and electrons in time $t$ is given by
\begin{equation}\label{LU}
    L_\text{thermal}=-\frac{dU}{dt}=-\frac{\mathcal{M}}{A m_\mathrm{p}}\bar{c}_v\frac{dT}{dt},
\end{equation}
where the mean specific heat is given by \eqref{cvg} with \eqref{cw}. However, we need to also take into account the latent heat which is released during the crystallization process. Assuming that it is given $\sim k_\mathrm{B} T$ (we will take the numerical coefficient equaled $1$ later on), the contribution to the luminosity resulting from this energy release is given by \cite{van1968crystallization}
\begin{equation}
    L_\text{latent}=k_\mathrm{B}T \frac{d(m_s/A m_\mathrm{p})}{dt},
\end{equation}
where $m_s$ is the amount of mass that is already crystallized. Let us write it as
\begin{align}\label{LL}
      L_\text{latent} 
    = k_\mathrm{B}T \frac{\mathcal{M}}{A m_\mathrm{p}} \frac{1}{\mathcal{M}}\dfrac{dm}{dr}\dfrac{dr}{d\rho}\dfrac{d\rho_s(T)}{dT}\frac{dT}{dt},
\end{align}
where we have introduced $\rho_s(T)$ as the density of the crystallized mass at a temperature $T$. 
\begin{question}{Exercise}
    Derive the equation \eqref{LL}.
\end{question}
The density of the crystallized mass  $\rho_s(T)$ is related to the ratio of Coulomb to thermal energy $\Gamma$ by \cite{koester1972outer}
\begin{equation}
    \Gamma=2.28\times10^5 \frac{Z^2}{A^{1/3}} \frac{\rho_s^{1/3}}{T}.
\end{equation}
The crystallization process starts when the above ratio reaches the critical value, it means when $\Gamma=\Gamma_m$.
\begin{question}{Exercise}
  Show that when the crystallization process is taken into account, one has
  \begin{equation}
        \frac{d\rho_s}{dT}= \frac{3\rho_s}{T}.
  \end{equation}
\end{question}
Using the above relation, the luminosity \eqref{LL} takes the following form
\begin{equation}\label{LL2}
    L_\text{latent} = 3\rho_s k_\mathrm{B} \frac{\mathcal{M}}{A m_\mathrm{p}} \frac{1}{\mathcal{M}}\frac{dm}{dr}\frac{dr}{d\rho}\frac{dT}{dt}.
\end{equation}
Let us note that $\frac{dm}{dr}$ and $\frac{d\rho}{dr}$ are taken at radius $r_*$ for which $\rho(r_*)=\rho_s(T)$ is satisfied.

Therefore, we have finally derived the cooling equation
\begin{equation}\label{LT}
    L = \frac{3k_\mathrm{B}\mathcal{M}}{A m_\mathrm{p}} \left(-\frac{\bar{c}_v}{3k_B} + \rho_s  \frac{1}{\mathcal{M}}\frac{dm}{dr}\frac{dr}{d\rho} \right)\frac{dT}{dt},
\end{equation}
in which the effect of modified gravity are given by the density profile $\rho(r)$ which is a solution of the modified equilibrium equation (or the modified LE equation). Similarly as in the case of brown dwarfs, such an equation allows to obtain the age of the object. It can be shown that white dwarfs in modified gravity one deals with younger objects than in Newtonian framework (that is, the stars cool down faster).

\section{Conclusions}

We have introduced a standard formalism which is used to provide an hydrostatical equilibrium of stellar and substellar objects in a framework of modified gravity. Moreover, we have also briefly discussed other equations, mainly related to the matter properties, which allow to describe internal processes happening in low-mass stars, brown dwarfs, giant planets as well as white dwarfs. As examples, we discussed light elements' ignition in the stellar cores, such as hydrogen and lithium burning. Regarding evolutionary phases, we demonstrated how to use the modified Lane-Emden formalism to examine cooling processes of brown and white dwarf stars.

There is however still a lot to do. We have just read about a few particular processes whose description was undertaken in the framework of modified gravity. As shown in \cite{baker}, we do not have too many tests of our models in the curvature regime of stars, which lies between the problematic one, that is, cosmological scales, and the well tested, that is, Solar System's and compact objects' one. This gap could
potentially hide the onset of corrections to General Relativity, urgely searched by the modified gravity community.

\begin{acknowledgement}
This work is supported by MICINN (Spain) {\it Ayuda Juan de la Cierva - incorporaci\'on} 2020 No. IJC2020-044751-I.
\end{acknowledgement}

\end{document}